\documentclass[fleqn]{article}

\newcommand{\titel}
{Periodic Single-Pass Instruction Sequences}

\usepackage{tikz,stmaryrd,amssymb,amsmath,amsthm,url}
\usetikzlibrary{shapes,arrows}

\newcommand{\rris}{\rrbracket_{\IS}}

\newtheorem{theorem}{Theorem}
  
\newtheorem{proposition}{Proposition}

\newtheorem{definition}{Definition}  
\theoremstyle{definition}
\newtheorem{example}{Example}

\newcommand{\FIS}{\ensuremath{K}}
\newcommand{\IS}{\ensuremath{K_r}}
\newcommand{\ISP}{\ensuremath{K_r^-}}

\newcommand{\PI}{\ensuremath{\mathcal U}}

\newcommand{\diamarrow}[1]{\ensuremath{\mathbin{   %
                  \setlength{\unitlength}{1.6ex}
                  \begin{picture}(3,2)
                  \put(.09,.1){$\langle~#1~\rangle$}
                  \put(.65,-.55){\vector(-1,-1){2.5}}
                  \put(2.35,-.55){\vector(1,-1){2.5}}
                  \end{picture}
                  }}}
\newcommand{\diamleftarrow}[1]{\ensuremath{\mathbin{ %
                  \setlength{\unitlength}{1.6ex}
                  \begin{picture}(3,2)
                  \put(.09,.1){$\langle~#1~\rangle$}
                  \put(.65,-.55){\vector(-1,-1){2.5}}
                  \put(2.35,-.55){\line(1,-1){2.5}}
                  \end{picture}
                  }}}
\newcommand{\prefa}[1]{\ensuremath{\mathbin{        %
                  \setlength{\unitlength}{1.6ex}
                  \begin{picture}(3,2)(-.54,.1)
                  \put(-0.3,.2){$\lbrack~#1~\rbrack$}
                  \put(.95,-.5){\line(0,-1){2.4}}
                  \end{picture}
                  }}}
\newcommand{\prefarrow}[1]{\ensuremath{\mathbin{    %
                  \setlength{\unitlength}{1.6ex}
                  \begin{picture}(3,2)(-.54,.1)
                  \put(-0.3,.2){$\lbrack~#1~\rbrack$}
                  \put(.95,-.5){\vector(0,-1){2.4}}
                  \end{picture}
                  }}}

\newcommand{\herhaal}[1]{\backslash\!\backslash\# #1}
\newcommand{\PGAr}{\ensuremath{E_{sc}}}
\newcommand{\PGAfr}{\ensuremath{E_{spc}}}

\newcommand{\di}{\mathsf{D}}
\newcommand{\st}{\mathsf{S}}

\newcommand{\lata}{\mathtt{pgla2pga}}

\newcommand{\Nat}{{\mathbb N}}
\newcommand{\Nplus}{{\mathbb N}_{>0}}

\newcommand{\tr}{{\mathtt{true}}}
\newcommand{\fa}{{\mathtt{false}}}

\title{\titel}

\author{
	Jan A.\ Bergstra \qquad
	Alban Ponse
\\[2mm]  {\small \begin{tabular}{c}
  Informatics Institute\\
  University of Amsterdam\\[2mm]
  \url{www.science.uva.nl/~janb/}\qquad 
	\url{www.science.uva.nl/~alban/}
	\end{tabular}}
}
\date{}

\begin{document}
\maketitle

\begin{abstract}
A program is a finite piece of data 
that produces
a (possibly infinite) sequence of primitive instructions. 
From scratch we develop a linear notation for sequential,
imperative programs,
using a familiar class of primitive 
instructions and so-called repeat instructions, a
particular type of control instructions. 
The resulting mathematical structure is a semigroup. 
We relate this set of programs to program algebra (PGA) and 
show that a particular subsemigroup is a carrier for
PGA by providing axioms for 
single-pass congruence, structural congruence, and
thread extraction. This subsemigroup characterizes 
periodic single-pass instruction sequences and  provides a
direct basis for PGA's toolset.
\\[1mm]
\end{abstract}

{\small\tableofcontents}

\section{Introduction}
\label{sec:intro}
Our starting point of view is that
a ``program'' is a finite piece of data for which the preferred or
natural interpretation (or meaning) is a sequence of
primitive instructions (SPI), and we say that a program
\emph{produces} a SPI. Primitive 
instructions comprise jump instructions, test instructions and 
basic instructions that upon execution
may alter some state; in the next section we introduce a  
set \PI\ of primitive instructions that we used in
previous research.

The execution
of a SPI is \emph{single-pass}: it starts with executing 
the first primitive instruction, and each
primitive instruction 
is dropped after it has been executed or jumped over.
This point of departure represents a most basic view of what 
a program constitutes: the syntactic denotation of
a SPI to be executed in single-pass mode.\footnote{In~\cite{BL02},
  SPIs are referred to as \emph{program objects}.}
A SPI can either be 
finite or infinite.

A very basic question is how to define
programs that produce the classes of SPIs we are interested in, 
adopting the point of view that a program itself is a finite 
sequence of instructions.
A first, very straightforward and simple
approach to this question is to start from constants for
primitive instructions and to adopt \emph{concatenation} as
an operation for composing 
programs: each primitive instruction is a program, and
if $P$ and $Q$ are programs,
then so is their concatenation $P;Q$. Furthermore, it is useful 
to postulate that concatenation is associative and (thus) to 
leave out brackets in repeated concatenations. 
This implies that programs built in this way represent the most 
simple set of programs that produce finite SPIs, and that in 
mathematical terms, this set constitutes a \emph{semigroup}.
We shall use the notation
\[K\]
for this very basic semigroup, where $K$ abbreviates
\emph{Kernel instruction sequence notation}.

However, in order to give an account of sequential, imperative
programming one needs programs that 
can produce certain \emph{infinite} SPIs 
(cf.~programs that define the finite-state control of
a Turing machine). 
An infinite SPI is \emph{periodic} if it can be produced by
a program of the form
\[u_1;\ldots;u_{n{+}k};\herhaal n\]
with $u_i\in\PI$, the set of primitive instructions, $n>0$ and $k\geq 0$, 
and with the \emph{repeat instruction} $\herhaal n$, which is
defined as follows: for $n$ a natural number larger than~0,
\[\herhaal n\]
prescribes to repeat the last preceding $n$ instructions. 
The repeat instruction, or briefly \emph{repeater}
is a \emph{control instruction} to be used for the definition of
a program that produces a periodic SPI. 
As an example with $u$ a primitive instruction, the program
$u;\herhaal 1$ (which consists of two instructions)
 produces the periodic SPI that consists
of an infinite number of $u$-instructions,
and the same SPI is 
produced by $u;u;\herhaal 1$ and by
many more  programs, for example by $u;u;\herhaal 2$ and 
$u;u;u;\herhaal 2$. 

In order to provide a setting for defining programs as finite
sequences of instructions, we define below two semigroups, where we write 
$S^+$ for the set of finite sequences with elements from alphabet $S$
and for which we (also) use  ``;'' as a separator:
\begin{itemize}
\item
The semigroup with domain $\PI^+$ and concatenation as its operation,
representing the finite SPIs \emph{and}
the finite programs  over \PI. As stated above, we use the name \FIS\
for this semigroup.
\item
The semigroup with domain $(\PI\cup\{\herhaal n\mid n\in\Nplus\})^+$
and concatenation as its operation; this semigroup will be used to 
represent periodic SPIs.
We use the name $\IS$ for this semigroup.
\end{itemize}

In Section~\ref{sec:2} we introduce the primitive instructions
we work with and for programs in \FIS\ and \IS,
we provide an axiomatization of
\emph{single-pass congruence}, the congruence that
identifies programs that produce identical SPIs.
We  discuss the fact that 
not each sequence of instructions in $\IS$ can be called a program. 
For example, the question whether the \IS-expression 
\[u;\herhaal 2\]
produces a SPI | and if so, which one | has no obvious answer.
We distinguish a subset of the domain of $\IS$ that
rules out this question and contains the \emph{\IS-programs} that produce 
all finite and periodic SPIs. 
In Section~\ref{sec:3} we discuss the behavior of
\IS-programs using \emph{thread algebra} and we define a
thread extraction operator that can be applied to \IS-programs.
For $\FIS$-programs and $\IS$-programs we provide axioms for 
 \emph{structural congruence}, a congruence that 
admits the unchaining of jump counters and preserves the
behavioral semantics of programs (thread extraction applied to 
structural congruent programs yields equal threads).
In Section~\ref{Conc}, we
relate our approach to PGA (program algebra, \cite{BL02}), 
which represents
the analysis of SPIs starting from a more
mathematically oriented design of a program notation for 
periodic SPIs (comprising a repetition \emph{operator} instead of
repeaters),
and to our program notation C~\cite{BP09}, a 
program notation design based on primitive
instructions that explicitly prescribe whether the orientation
of the execution order is left-to-right or vice versa.

\section{\FIS-basics}
\label{sec:2}
In this section we formally define our set of primitive instructions 
(taken from \cite{BL02}) and the semigroups \FIS\ and \IS. Then we discuss
canonical forms as a preferred form of representation of 
\IS-expressions and define  \emph{\IS-programs}.

\subsection{Primitive instructions, \FIS\ and \IS}
Let $A$ be a set of constants and 
write $\Nplus$ for $\Nat\setminus\{0\}$, where $\Nat$ represents the
natural numbers.

\begin{definition}
\label{def:KKr}
\textbf{\FIS-expressions}, also called \textbf{\FIS-programs}, 
are defined by the following grammar, where 
$\mathtt{a}\in A$ and $k\in\Nat$:
\[P ::= \mathtt{a}\mid +\mathtt{a}\mid -\mathtt{a}
\mid \#k\mid\;! \mid P;P\]
and where the operation $;$ is called \textbf{concatenation}.

\textbf{\IS-expressions} are defined by the following grammar, where 
$\mathtt{a}\in A$ and $k\in\Nat,~n\in\Nplus$:
\[P ::= \mathtt{a}\mid +\mathtt{a}\mid -\mathtt{a}
\mid \#k\mid\;! \mid \herhaal n\mid P;P\]
(\textbf{\IS-programs} are defined in Definition~\ref{def:Kr}).
\end{definition}

Let $\mathtt a\in A$ and $k\in\Nat$. 
Then each of 
$\mathtt{a}, +\mathtt{a}, -\mathtt{a}, \#k, !$ is called a  
\emph{primitive instruction} and primitive instructions occurring
in \FIS-expressions can be explained as follows:

\begin{itemize}
\item
A \emph{basic instruction} $\mathtt a\in A$ prescribes an atomic
piece of behavior that is considered
indivisible and executable in finite time. After completion of its 
execution, it prescribes to execute the next instruction (if available).
One can consider various specific instances of $A$ and we mention here
the set of \emph{molecular programming primitives}, see, e.g.,~\cite{BB02}.
\item A basic instruction can be turned into a 
\emph{test instruction} by prefixing it with either the symbol $+$ 
(positive test instruction) or with  the symbol $-$ (negative test instruction), 
thus typically $\mathtt{+a}$, $\mathtt{-b}$ etc. 
Test instructions control subsequent 
execution via the result of their execution, which is
a Boolean reply that may depend on the execution state:\footnote{
Upon reply $\tr$, a positive test instruction prescribes
to execute the next instruction (if available) and upon reply
$\fa$ it prescribes to skip the next instruction and to proceed execution
with the instruction thereafter. A negative test instruction has the same 
effect, but with the role of the replies reversed.}
we explain this in detail in Section~\ref{sec:TA}.

\item
A next kind of primitive instruction is the
\emph{jump instruction} $\#k$ where $k\in\Nat$: this instruction
prescribes to jump $k$ primitive
instructions ahead (if possible; otherwise deadlock
occurs) and generates no observable
behavior. The special case  $\#0$ prescribes deadlock.
\item The
\emph{termination instruction} $!$ prescribes successful
termination, an event that is taken to be observable.
\end{itemize}
We write \PI\ for the set of primitive instructions and we 
shall use $u, u_1, u_2,..., v, v_1,v_2,...$
as typical variables for elements in \PI.
We define each element of \PI\ to be a SPI (Sequence of 
Primitive Instructions). 

\emph{Finite} SPIs are produced by \FIS-expressions (see Definition~\ref{def:KKr}).
We take concatenation to be an \emph{associative} operator 
and leave out brackets in repeated concatenations, so we simply
write 
\[u_1;u_2;\ldots;u_n\]
for the \FIS-expression built up from the primitive instructions 
$u_1, ... ,u_n$ by $n{-}1$ repeated concatenations.  
Thus \FIS\ is the free semigroup with generators from \PI.

Finally, the non-primitive \emph{repeat instruction} $\herhaal n$, where
$n\in\Nplus$, prescribes to repeat the last preceding $n$ instructions.
Repeat instructions are also called \emph{repeaters}.
So-called \emph{periodic} SPIs are produced by \IS-expressions (see
Definition~\ref{def:KKr}). Again, we 
take concatenation to be an \emph{associative} operator and
leave out brackets in repeated concatenations,
thus \IS\ is the free semigroup generated by $\PI\cup\{\herhaal n\mid n\in\Nat\}$.
By definition, \FIS\ is a subsemigroup of \IS.

\subsection{Single-pass congruence and first canonical forms}
In this section we define 
\emph{single-pass congruence}, the congruence
that characterizes extensional equality of SPIs, i.e.,
the equality defined by having
the same primitive instruction 
at each position in the SPI that is produced.\footnote{Although a
  bit long, \emph{primitive instruction sequence congruence} 
  would also be an adequate name.}

For \FIS-expressions, thus \IS-expressions not containing repeaters,
single-pass congruence boils
down to the associativity of concatenation.

\begin{table}
\centering
\hrule
\begin{align}
\label{P2}
(u_1;\ldots;u_{n})^{m};\herhaal mn
&=
u_1;\ldots;u_{n};\herhaal n
\\[1mm]
\label{P3}
\herhaal n;X&=\herhaal n\\[1mm]
\label{P4}
u_1;\ldots;u_{m};v_1;\ldots;
v_{n};u_1;\ldots;u_{m};\herhaal m{+}n
&=
u_1;\ldots;u_{m};v_{1};\ldots;v_{n};
\herhaal m{+}n
\end{align}
\hrule
\caption{The axiom set \PGAfr\ for single-pass congruence
on \IS-expressions, 
where $m,n\in\Nplus$ and $u_i,v_j\in\PI$}
\label{tab:SI1}
\end{table}

Define for $n>0$ and $X$ an \IS-expression, 
$X^{n+1}=X;X^n$ and $X^1=X$. 
Single-pass congruence for \IS-expressions 
is axiomatized by the axiom schemes
\eqref{P2}--\eqref{P4} in Table~\ref{tab:SI1}
and equational logic, and we write
\[\PGAfr\] 
for this proof system. 
Although equations~\eqref{P2}--\eqref{P4} are in fact
\emph{schemes} 
in $m,n\in\Nplus$, we further refer to these as ``axioms''. 
Whenever two \IS-expressions $X$ and $Y$ are single-pass
congruent, this is written \[X=_{spc}Y,\]
and the subscript ${spc}$ will be dropped if no confusion can arise.

\begin{proposition}
The unfolding property
\[u_1;\ldots;u_n;\herhaal n=
(u_1;\ldots;u_n)^2;\herhaal n\]
follows from $\PGAfr$.
\end{proposition}

\begin{proof}
\begin{align*}
\PGAfr\vdash u_1;\ldots;u_n;\herhaal n
&=(u_1;\ldots;u_n)^2;\herhaal 2n
&&\text{by~\eqref{P2}}\\
&=u_1;\ldots;u_n;(u_1;\ldots;u_n)^2;\herhaal 2n
&&\text{by~\eqref{P4}}\\
&=u_1;\ldots;u_n;u_1;\ldots;u_n;\herhaal n
&&\text{by~\eqref{P2}}\\
&=(u_1;\ldots;u_n)^2;\herhaal n.
\end{align*}
\end{proof}

In \PGAfr, axiom~\eqref{P3}
implies that each expression in $\IS$
can be equated to one that
contains \emph{at most} one repeat instruction.
This leads to the following preferred
representation of \IS-expressions.

\begin{definition}
\label{cfr}
A \IS-expression is a \textbf{first canonical form}
if it is of the form
\[u_1;\ldots;u_{n} \quad\text{or}\quad 
u_1;\ldots;u_k;\herhaal n\]
with $u_i\in\PI$, $k\in\Nat$ and $n\in\Nplus$,
where $u_1;\ldots;u_0;$ represents the empty sequence.
\end{definition}

For each \IS-expression, its {first canonical form}
is obtained by applying axiom~\eqref{P3}
to the leftmost occurring repeater if present, and otherwise it is
that expression itself.

Not all \IS-expressions have an intuitive meaning.
For example, 
\[
\mathtt a;\herhaal 2 \quad\text{and}\quad
\#7;+\mathtt a;\herhaal 5
\]
illustrate this situation. Note that such
first canonical forms
can not be rewritten using 
any of the axioms~\eqref{P2}--\eqref{P4} in Table~\ref{tab:SI1}.
As a consequence, single-pass congruence is not a meaningful
notion for such first canonical forms and in the next section we
will exclude such \IS-expressions.

\subsection{\IS-programs and their first canonical forms}
Let $\ISP$ stand for the subset of \IS-expressions whose
first canonical form has
the property that the repeat instruction $\herhaal n$ (if present)
is preceded by at least $n$ primitive instructions. 
In fact, \ISP\ is a subsemigroup of \IS: if $X,Y\in\ISP$, then
$X;Y\in\ISP$. In the following definition,
we refine the notion of a first canonical form.

\begin{definition}
\label{def:minimal}
Let $u_1;\ldots;u_{k}$ with $k\geq 1$ be a SPI (thus all $u_i$ are
primitive instructions). 
Then $u_1;\ldots;u_{k}$ is a 
\textbf{first canonical $\IS$-form}. This first 
canonical $\IS$-form is 
\textbf{minimal} by definition.

The \IS-expression
$u_1;\ldots;u_k;\herhaal n$
is a \textbf{first canonical $\IS$-form}
if $0<n\leq k$. This first canonical $\IS$-form
is \textbf{minimal} if its repeating part 
$u_{k-n+1};\ldots;u_k$
can not be made smaller
with axiom~\eqref{P2}, and its non-repeating part
 $u_1;\ldots;u_{k{-}n}$ can not be made smaller
with axiom~\eqref{P4}.
\end{definition}

Two examples, where the right-hand sides are minimal first 
canonical $\IS$-forms: 
\begin{align*}
+\mathtt a; -\mathtt b;\#4;-\mathtt b;\#4;\herhaal 4
&=_{spc}+\mathtt a; -\mathtt b;\#4;\herhaal 2,\\
-\mathtt a; +\mathtt c;\#4;+\mathtt c;\herhaal 2;+\mathtt b
&=_{spc}-\mathtt a; +\mathtt c;\#4;\herhaal 2.
\end{align*}

\begin{definition}
\label{def:Kr}
Elements in \ISP\ are referred to as \textbf{\IS-programs}.
\end{definition}

Recall that
two $\IS$-programs $P$ and $Q$ are single-pass
congruent if, and only if,
\[\PGAfr\vdash P=Q.\]
Single-pass congruence for $\IS$-programs
is captured by the next result. 

\begin{theorem}
\label{thm:1}
Single-pass congruence of \IS-programs is decidable.
\end{theorem}

\begin{proof}
Assume $P$ and $Q$ are two \IS-programs that
denote identical SPIs. If both programs
do not contain repeat instructions, they are syntactically
identical, apart from the possible use of brackets, which is 
then captured by the associativity of concatenation, which we adopted 
throughout this paper. 
In the other case,
application of axiom~\eqref{P3} yields first 
canonical $\IS$-forms and these expressions still
denote the same SPI. With axiom~\eqref{P4} the 
non-repeating parts of the two \ISP-expressions (if present) 
can be made as 
short as possible, so these should be identical for both 
expressions. Removal of these non-repeating parts yields
two expressions of the form
$u_1;\ldots;u_n;\herhaal n$ and $v_1;\ldots;v_m;\herhaal m$
that denote identical SPIs. With axiom~\eqref{P2} one then derives
\begin{align*}
\PGAfr\vdash u_1;\ldots;u_n;\herhaal n
&=(u_1;\ldots;u_n)^m;\herhaal nm\\
&=(v_1;\ldots;v_m)^n;\herhaal nm\\
&=v_1;\ldots;v_m;\herhaal m.
\end{align*}

Of course, the values $n$ and $m$ in 
these repeating parts 
can be effectively minimized with 
axiom~\eqref{P2}. Then 
single-pass congruence coincides with the syntactic equality 
of both minimal first canonical \IS-forms, which
immediately implies the mentioned decidability.
\end{proof}
Without loss of generality, we further only consider
\IS-programs that contain at most
one repeat instruction.
\section{Execution of \IS-programs}
\label{sec:3}
We briefly discuss Thread Algebra (cf. \cite{PZ06}), 
earlier described in e.g. \cite{BB03,BL02}.
For basic information on thread algebra we refer 
to~\cite{BBP05,PZ06}; more advanced matters, such as an operational
semantics for thread algebra, are discussed
in~\cite{BM07}.

\subsection{Thread algebra}
\label{sec:TA}
\emph{Threads} model the execution of SPIs.
In order to define threads, we consider the set 
$A$ of basic instructions also as a set of so-called \emph{actions}
that model the execution of
basic and test instructions, where it is assumed 
that execution of the action $\mathtt a$ yields a Boolean 
reply $\tr$ or $\fa$. 
Finite threads are defined inductively in the following way:
\begin{align*}
&\st&&\text{\emph{stop}, the termination thread,}\\[1mm]
&\di&&\text{\emph{inaction} or \emph{deadlock}, the inactive thread},\\[1mm]
&P\unlhd\mathtt a \unrhd Q&&
\text{the \emph{postconditional
composition} $\_\unlhd\mathtt a \unrhd\_$ of finite threads $P$ and $Q$,}\\
&&&\text{where $\mathtt a\in A$.}
\end{align*}
The behavior of the thread
$P\unlhd\mathtt a \unrhd Q$ starts with the \emph{action} $\mathtt
a$ and continues as $P$ upon reply $\tr$ to $\mathtt a$, and as $Q$
upon reply $\fa$. Note that finite threads always end in $\st$ or
$\di$.
We use \emph{action prefix} $\mathtt a \circ P$ as an abbreviation for
$P\unlhd\mathtt a \unrhd P$ and take $\circ$ to bind strongest.

A so-called \emph{regular} thread over $A$ is
a finite-state thread in which infinite paths can occur 
(so, finite threads  form a special subset of regular threads).
Each regular thread can be defined by a finite number of recursive 
equations.
As a first example, consider the regular thread $Q$ defined by
\begin{align*}
Q&=\mathtt a\circ R,\\
R&=\mathtt c\circ R\unlhd \mathtt b\unrhd(\st\unlhd \mathtt d\unrhd Q).
\end{align*}
This regular thread $Q$ can be depicted in the following way:
{\small
\tikzstyle{decision} = [diamond, aspect=1.8, draw, fill=blue!10, 
    text width=2.8em, text badly centered, node distance=2cm, inner sep=0pt]
\tikzstyle{block} = [rectangle, draw, fill=blue!10, 
    text width=2em, text centered, minimum height=1.8em]
\tikzstyle{block2} = [ellipse, draw, fill=blue!10, 
    text width=1.4em, text centered, rounded corners, minimum height=1.4em]
\tikzstyle{line} = [draw, thick, -latex]
\tikzstyle{cloud} = [draw, ellipse,fill=red!10, node distance=2cm,
    minimum height=2em]
\tikzstyle{leeg} = [node distance=2cm, inner sep=0pt,
    minimum height=0em]
\\[6mm]
\phantom{\hspace{1cm}}
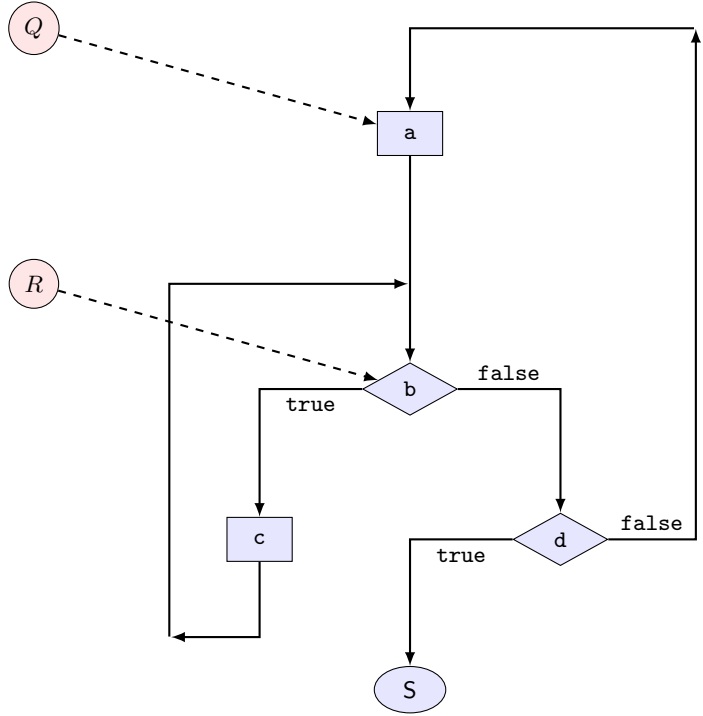
\begin{tikzpicture}[node distance = 1cm, auto]
    \node [leeg] (L1) {};
    \node [block, below of=L1, node distance=1.4cm] (a) {$\mathtt a$};
    \node [leeg, below of=a] (L2) {};
    \node [decision, below of=a, node distance=3.4cm] (b) {$\mathtt b$};
    \node [cloud, left of=L1, node distance=5cm] (Q) {$Q$};
    \node [leeg, below of=b] (L3) {};
    \node [cloud, left of=L2, node distance=5cm] (R) {$R$};
    \node [block, left of=L3, node distance=2cm] (c) {$\mathtt c$};
    \node [decision, right of=L3, node distance=2cm] (d) {$\mathtt d$};
    \node [block2, below of=L3, node distance=2cm] (S) {$\st$};
    \node [leeg, below of=c, node distance=1.3cm] (L4) {};
    \node [leeg, below of=d, node distance=2.8cm] (L6) {};
    \node [leeg, left of=L4, node distance=1.2cm] (R1) {};
    \node [leeg, right of=L1, node distance=3.8cm] (R0) {};
    \path [line] (a) -- (b);
    \path [line] (b) -| node  [near start] {$\tr$}(c);
    \path [line] (b) -| node  [near start] {$\fa$}(d);
    \path [line] (d) -| node  [near start] {$\tr$}(S);
    \path [line,dashed] (Q) -- (a);
    \path [line,dashed] (R) -- (b);
    \path [line] (d) -| node [near start]{$\fa$}(R0);
    \path [line] (R0) -| (a);
    \path [line] (R1) |- (L2);
    \path [line] (c) |- (R1);
\end{tikzpicture}
\\[-2mm]
}
\indent
Each regular thread can be specified using a so-called linear recursive specification.

\begin{definition}
\label{def:linear}
A \textbf{linear recursive specification} is a set of equations
\[\{P_i=t_i(\overline P)\mid i=0,...,n\}\]
with $t_i(\overline P)$  of the form $\st,\di$, or 
$P_{i_1}\unlhd \mathtt a_i\unrhd P_{i_2}$, where $\mathtt a_i\in A$ and $i_1,i_2\leq n$. 
\end{definition}

For the example above, we find $Q=P_0$ for $P_0$ defined by the following linear equations:
\begin{align*}
P_0&= P_1\unlhd \mathtt a\unrhd P_1,\\
P_1&=P_2\unlhd\mathtt b\unrhd P_3,\\
P_2&=P_1\unlhd \mathtt c\unrhd P_1,\\
P_3&=P_4\unlhd \mathtt d\unrhd P_0,\\
P_4&=\st.
\end{align*}
In the next section we explain in what way $\IS$-programs
define regular threads.

\subsection{Behavioral semantics for $\IS$-programs: threads}
As mentioned before, the execution
of a SPI is \emph{single-pass}: it starts with the 
first
instruction, and each instruction is dropped after it has been
executed or jumped over. In this section we explain
the precise \emph{meaning} of primitive 
instructions and $\IS$-programs in terms of their execution.

\begin{table}
\centering
\hrule
\begin{align*}
&\text{Let $X= u_1;\ldots;u_{n{+}k};\herhaal n$,~~then $\llbracket X\rris=|1,X|$, where}\\[3mm]
|j, X|&=|j{-}n, X| \quad\text{ if }j>n{+}k,\\[1mm]
|j, X|&=\st \quad\text{ if } u_j=!,\\[1mm]
|j, X|&=a\circ |j{+}1, X| \quad\text{ if } u_j = \mathtt a,\\[1mm]
|j, X|&=|j{+}1, X| \unlhd a\unrhd |j{+}2, X|
\quad\text{ if }u_j = +\mathtt a,\\[1mm]
|j, X|&=|j{+}2, X| \unlhd a\unrhd |j{+}1, X|
  \quad\text{ if }u_j = -\mathtt a,\\[1mm]
|j, X|&=\di \quad\text{  if }u_j = \#0,\\[1mm]
|j, X|&=|j{+}k{+}1, X| \quad\text{  if }u_j = \#k{+}1.
\end{align*}
\hrule
\caption{Equations for thread extraction on $\ISP$,
where $u_i\in\PI$, $k\in\Nat$ and $j,n\in\Nplus$}
\label{tab:te}
\end{table}

Let $X$ be a $\IS$-program of the form
\[X=u_1;\ldots;u_{n{+}k};\herhaal n,\]
thus $X$ is a first canonical 
\IS-form. In  Table~\ref{tab:te} we define the 
\emph{thread extraction} of $X$, 
notation 
\[\llbracket X\rris,\]
where the auxiliary function
\(|j,X|\)
models the thread extraction of program $X$ when started at its  
$j$'th instruction.
In the general case of a $\IS$-program $X$, its 
thread extraction is defined by
\[\llbracket X\rris=\llbracket X;\#0;\herhaal 1\rris,\]
thus the SPI produced by $X$ that | if it is finite | is extended 
with an 
infinite number of $\#0$-instructions.
Because each $\IS$-program of the form $X;\#0;\herhaal 1$
can be converted to a first canonical form 
$u_1;\ldots;u_{n{+}k};\herhaal n$, the equations in 
Table~\ref{tab:te} match all possible cases:
\begin{itemize}
\item Repeaters and the termination instruction $!$ are dealt with
in the first two equations for $|j,X|$. Observe
that termination must always be explicitly defined using $!$.
\item
A basic or test
instruction yields the equally named action in a post
conditional composition.
In the case of a positive
test instruction $+\mathtt a$, the reply $\tr$ to the associated
action $\mathtt a$ prescribes to continue with
the next instruction and the reply $\fa$ prescribes
to skip the next instruction
and to continue with the instruction at the position thereafter; 
for the execution of a negative
test instruction $-\mathtt a$, subsequent execution
is prescribed by the complementary replies.
If there is no next instruction to be executed,
deadlock follows. 

\item A $\#0$-instruction yields deadlock upon execution, and 
a jump instruction $\#k{+}1$ shifts $|j,X|$ to $|j{+}k{+}1,X|$.
\end{itemize}

A first, very simple example is the regular thread
obtained by thread extraction of 
the $\IS$-program $+\mathtt a;\herhaal 1$. We 
find that this program prescribes the execution of an 
infinite sequence of $\mathtt a$-actions:
\begin{align*}
\llbracket +\mathtt a;\herhaal 1\rris
=|1,+\mathtt a;\herhaal 1|
&=|2,+\mathtt a;\herhaal 1|\unlhd \mathtt a\unrhd |3,+\mathtt a;\herhaal 1|\\
&=|1,+\mathtt a;\herhaal 1|\unlhd \mathtt a\unrhd |2,+\mathtt a;\herhaal 1|\\
&=\mathtt a\circ |1,a;\herhaal 1|,
\end{align*}
and thus the regular thread captured by the single recursive
equation
\begin{equation}
\label{eq:t2}
\tag{$e_1$}
P=\mathtt a\circ P
\end{equation}
and we may write $\llbracket +\mathtt
a;\herhaal 1\rris=P$ for
$P$ defined as in equation~\eqref{eq:t2}.

The equations
in Table~\ref{tab:te} need not immediately yield a   
{regular} thread for each $\IS$-program:
it can be the case that these equations can be
consecutively applied without yielding any action, as for example
for the program $X=\#4;\mathtt a;\herhaal 2$, for which we derive
\begin{align*}
\llbracket X\rris=|1,X|
&=|5,X|\\
&=|3,X|\\
&=|1,X|.
\end{align*}
In such 
cases we define the extracted behavior to be $\di$, and
with this \emph{default-rule} for thread extraction, 
each $\IS$-program defines a regular thread.

\begin{example}
\label{ex1}\it
Let $X=+\mathtt a; \#0;+\mathtt b;\#4;-\mathtt c;\#0;\herhaal 4$.
We show that $\llbracket X\rris=
\di\unlhd\mathtt a \unrhd P$ with $P$ defined by the recursive 
equation
$P=\di\unlhd\mathtt b\unrhd (P\unlhd\mathtt c\unrhd \di)$.

Let $Y=+\mathtt a; \#0;+\mathtt b;\#4;-\mathtt c;\herhaal 4$, then
$X=_{spc}Y$
by axiom~\eqref{P4} and hence 
$\llbracket X\rris=\llbracket Y\rris$.
We first derive an intermediate result:
\begin{align*}
|4,Y|&=|8,Y|\\
&=|4,Y|,
\end{align*}
so by the default-rule, $|4,Y|=\di$. Finally, we derive
\begin{align*}
\llbracket Y\rris&=|1,Y|\\
&=|2,Y|\unlhd a\unrhd |3,Y|\\
&=\di\unlhd a\unrhd P,\\[2mm]
\text{where }P&=|3,Y|\\
&=|4,Y|\unlhd b\unrhd |5,Y|\\
&=\di\unlhd b\unrhd (|7,Y|\unlhd c\unrhd |6,Y|)\\
&=\di\unlhd b\unrhd (|3,Y|\unlhd c\unrhd |2,Y|)\\
&=\di\unlhd b\unrhd (P\unlhd c\unrhd \di).
\end{align*}
\end{example}

Conversely,
each regular thread over $A$
can be specified (programmed) by a $\IS$-program,
as we will discuss in Section~\ref{XX}.
For example, the regular thread $Q$ that was discussed 
above and that was specified by the equations
\begin{align*}
Q&=\mathtt a\circ R,\\
R&=\mathtt c\circ R\unlhd \mathtt b\unrhd(\st\unlhd \mathtt d\unrhd Q),
\end{align*}
satisfies
$Q=\mathtt{\llbracket a;+b;\#2;\#3;c;\#4;+d;!;\herhaal 8
\rris}$.

To conclude this section, we  mention the fact that 
in terms of execution behavior, certain different regular threads
should be considered equal, e.g.,
\[\llbracket \mathtt a;\herhaal 1\rris
\quad\text{and}\quad
\llbracket +\mathtt a;\mathtt a;\herhaal 2\rris\]
because both perform repeatedly the action $\mathtt a$ and are
thus behaviorally equivalent. A formal way to prove this
behavioral equivalence is discussed in~\cite{BL02} 
(and summarized in~\cite{PZ06}) and is considered outside
the scope of this paper.
Finally, observe that behavioral equivalence of $\IS$-programs, 
say $\equiv_{be}$, is not a congruence: although 
$\mathtt a\equiv_{be}+\mathtt a$ because both define the thread
$\mathtt a\circ\di$, we
find $\mathtt a;!\not\equiv_{be}+\mathtt a;!$ because 
$\mathtt a\circ\st \ne \st\unlhd\mathtt a\unrhd \di$.

\subsection{$\IS$-programs, second canonical forms and thread extraction}
One can change the jump counters in $\IS$-programs while preserving
execution behavior, for example
\[+\mathtt a;\#2;+\mathtt b;\#2;c;d;e\quad
\text{and}\quad
+\mathtt a;\#5;+\mathtt b;\#2;c;d;e\]
execute apart from their jump counters the same instructions and 
their 
thread extraction yields identical threads. The crucial difference
between these programs is that the rightmost program contains no 
\emph{chained jumps}.
In Table~\ref{pgar2} we introduce the axiom
schemes~\eqref{P5}--\eqref{P8} for the unchaining of jump
instructions and we write 
\[\PGAr\]
for the extension of $\PGAfr$ with these axiom
schemes.
Although~\eqref{P2}--\eqref{P8} are
axiom schemes 
in $m, n\in\Nplus$ and $k,\ell\in \Nat$, we shall refer to 
all of these as ``axioms''. The congruence defined by \PGAr\ is called
\emph{structural congruence}, and whenever two $\IS$-programs 
$X$ and $Y$
are structurally congruent, this is written
\[X =_{sc} Y.\]
Note that 
first canonical forms not in \ISP\ (thus, with a repeat counter
that is too large) can not be rewritten using 
any of the axioms~\eqref{P2}--\eqref{P8} in Table~\ref{pgar2}
that contain repeaters.
As a consequence, structural congruence is not a meaningful
notion for such first canonical forms.

\begin{table}
\centering
\hrule
\begin{align}
\tag{\ref{P2}}
(u_1;\ldots;u_{n})^{m};\herhaal mn
&=
u_1;\ldots;u_{n};\herhaal n
\\[1mm]
\tag{\ref{P3}}
\herhaal n;X&=\herhaal n\\[1mm]
\tag{\ref{P4}}
u_1;\ldots;u_{m};v_1;\ldots;
v_{n};u_1;\ldots;u_{m};\herhaal m{+}n
&=
u_1;\ldots;u_{m};v_{1};\ldots;v_{n};
\herhaal m{+}n
\\[4mm]
\label{P5}
\tag{4}
\#k{+}1;u_{1};\ldots;u_{k};\# 0
&= 
\#0;u_{1};\ldots;u_{k};\# 0
\\[1mm]
\label{P6}
\tag{5}
\#k{+}1{+}n;u_{1};\ldots;u_{k};\# n
&= 
\#k{+}1;u_{1};\ldots;u_{k};\# n
\\[1mm]
\label{P7}
\tag{6}
\#k{+}1{+}\ell;u_{1};\ldots;u_{k};\herhaal k{+}1
&=
\#\ell;u_{1};\ldots;u_{k};\herhaal k{+}1\\[1mm]
\label{P8}
\tag{7}
\#k{+}1{+}\ell{+}n;u_{1};\ldots;u_{k};v_{1};\ldots;v_{n};\herhaal n
&=\#k{+}1{+}\ell;u_{1};\ldots;u_{k};v_{1};\ldots;v_{n};\herhaal n\end{align}
\hrule
\caption{The axiom set \PGAr\ for structural congruence on SPIs, 
where $k,\ell\in\Nat$, 
$m,n\in\Nplus$, $u_i,v_j\in\PI$, and $u_{1};\ldots;u_{0};$
represents the empty sequence}
\label{pgar2}
\end{table}

\begin{definition}
\label{cf}
A \textbf{second canonical $\IS$-form} is a first 
canonical $\IS$-form
in which no chained jumps occur, and in the case of
$u_1;\ldots;u_m;\herhaal n$, in which
all jumps to $u_{m{-}n{+}1},\ldots,u_m$
are minimized using axioms~\eqref{P7} and \eqref{P8}.
\end{definition}

Two typical examples, where the $\IS$-programs in the 
right-hand column are second canonical 
$\IS$-forms (and those in the left-hand column are not):
\begin{align*}
\#1;\herhaal 1&=_{sc}\#0;\herhaal 1,\\
+\mathtt a; \#2;+\mathtt b;\#2;-\mathtt c;\#4;\herhaal 4&=_{sc}
+\mathtt a; \#0;+\mathtt b;\#0;-\mathtt c;\#0;\herhaal 4\\
&=_{sc}
+\mathtt a; \#0;+\mathtt b;\#0;-\mathtt c;\herhaal 4.
\end{align*}
The first example is an instance of axiom~\eqref{P7} 
($k=\ell=0$). The last example also provides
second canonical \IS-forms for the $\IS$-program
considered in Example~\ref{ex1}.

It is easily seen that in \PGAr,
second canonical $\IS$-forms have a unique minimal
representation in terms of their number of
instructions and we have the following result.

\begin{theorem}
Structural congruence of $\IS$-programs
is decidable, and two $\IS$-programs $P$ and $Q$ are structurally
congruent if, and only if,
\[\PGAr\vdash P=Q.\]
\end{theorem}

\begin{proof}
By the proof of Theorem~\ref{thm:1} it suffices to consider minimal
first canonical $\IS$-forms, and it
is trivial to convert such a form
to a second canonical $\IS$-form. Minimization of the length
of the
repeating part with~\eqref{P2} then yields a unique $\IS$-program
(cf.~the last example above).
\end{proof}

Thread extraction is more
straightforward when applied to second canonical $\IS$-forms:
structural congruent $\IS$-programs define
identical threads and because the infinite chaining of jumps is
excluded, the rules in Table~\ref{tab:te} are then complete and
there is no more need for the \emph{default-rule} (that stated that 
whenever the 
equations do not yield any action, the resulting behavior is $\di$). 

\section{Discussion and conclusions}
\label{Conc}
Our main motivation to undertake the current research is 
that in the setting of program algebra (PGA), the notion of a `program notation' as
defined in~\cite{BL02} should be strengthened, and
we return to this question in Section~\ref{YY}.
Program algebra was introduced 
as a general approach to model and analyze the notion of a 
sequential, imperative program in the form of a 
rather `non-formal' and theoretical style. An algebra of these 
programs named PGA is used as the carrier for a further 
development of this matter, and the syntax of PGA serves as a very simple and
basic program notation, underlying many other program notations.
In Section~\ref{XX} we relate our approach to PGA.

In Section~\ref{YY} we conclude the paper
with a consideration about PGLA, an earlier account
of the semigroup \IS\ that was proposed as a machine-readable 
version of PGA and that underlies the current toolset for 
PGA~\cite{Diertens}: $\IS$-expressions are precisely the programs that
can be processed by this toolset.

PGA can be viewed as a theory of instruction sequences with 
our subsemigroup $\ISP$ or PGLA as one of its many representations.
Unfortunately, we have not been able to identify any pre-existing theory
by other authors to which this work can be related in a convincing
manner.
The phrase \emph{instruction sequence} 
seems not to play a clear role in the theory of programming, and
the software engineering literature at large features many uses of this
phrase, but only in a casual setting. 

\subsection{Program algebra}
\label{XX}
PGA was set up in a very similar way as \IS, with the only 
difference that instead of repeat instructions, a unary 
operator called
\emph{repetition} is used. The notation for this operator is
\[(\_)^\omega\]
and its relation with \IS\ is captured by the equation scheme 
\begin{equation}
\label{eq:IS}
\tag{$e_2$}
u_1;\ldots;u_k;\herhaal k=(u_1;\ldots;u_k)^\omega\quad\text{for $k\in\Nplus$},
\end{equation}
where $u_1,...,u_k\in \PI$, the set of
primitive instructions that PGA and \IS\ are based on.

The associativity of concatenation and all further axiomatizations 
discussed previously, that is, the axiomatizations for
single-pass congruence and structural congruence for \IS\ are the direct
counterparts of those provided for PGA
when equation scheme~\eqref{eq:IS} is applied, and the same holds for the 
equations that define thread extraction. 
From a mathematical point of view, 
all such axioms and equations
formulated in the setting of PGA
are more elegant. The 
axiomatization of single-pass congruence in PGA is indeed
so simple that it can be easily remembered by heart:
\begin{align*}
(X;Y);Z&=X;(Y;Z)&
\mathrm{(PGA1)}\\[1mm]
(X^n)^\omega&=X^\omega
&\mathrm{(PGA2)}\\[1mm]
X^\omega;Y&=X^\omega
&\mathrm{(PGA3)}\\[1mm]
(X;Y)^\omega&=X;(Y;X)^\omega&\mathrm{(PGA4)}
\end{align*}

However, in terms of a \emph{program notation} for finite or periodic SPIs, there is something to be said \emph{against} 
PGA: its notation is not conforming to ASCII and 
exploits a scope-dependent unary operator $(\_)^\omega$.

Because of the immediate correspondence between $\IS$-programs
and PGA-programs as characterized by equation scheme~\eqref{eq:IS},
many PGA-results also hold for $\IS$-programs.
For example, each regular thread over $A$
can be specified (programmed) by a $\IS$-program: assume
a regular thread $P_0$ is given by the linear recursive
specification
\[\{P_i=t_i(\overline P)\mid i=0,...,n\}\]
with $t_i(\overline P)$  of the form $\st,\di$, or 
$P_{i_1}\unlhd \mathtt a_i\unrhd P_{i_2}$, 
where $\mathtt a_i\in A$ and $i_1,i_2\leq n$.
Then 
\[P_0= {\llbracket \sigma_0(P_0);\ldots;\sigma_n(P_n);
\herhaal 3n+3\rris},\]
where
\[\sigma_i(P_i)=\begin{cases}
!;\#0^2&\text{if }t_i=\st,\\
\#0^3&\text{if }t_i=\di,\\
+\mathtt a;\#f(n,i_1,i_2);\#g(n,i_1,i_2)&\text{if }
t_i=P_{i_1}\unlhd \mathtt a_i\unrhd P_{i_2},\end{cases}
\]
for appropriate target functions $f$ and $g$. For example,
the regular thread $Q$ discussed 
in Section~\ref{sec:TA} and specified by the equations
\begin{align*}
Q&=\mathtt a\circ R,\\
R&=\mathtt c\circ R\unlhd \mathtt b\unrhd(\st\unlhd \mathtt d\unrhd Q),
\end{align*}
and thus by $P_0$ in the linear recursive specification (see Definition~\ref{def:linear}) that consists of the linear equations
\begin{align*}
P_0&= P_1\unlhd \mathtt a\unrhd P_1,\\
P_1&=P_2\unlhd\mathtt b\unrhd P_3,\\
P_2&=P_1\unlhd \mathtt c\unrhd P_1,\\
P_3&=P_4\unlhd \mathtt d\unrhd P_0,\\
P_4&=\st,
\end{align*}
satisfies
\begin{align*}
Q=P_0=\llbracket&\mathtt{ +a;\# 2;\# 1;}
\\
&\mathtt{+b;\#2;\#4;}\\
&\mathtt{+c;\#11;\# 10;}\\
&\mathtt{+d;\# 2;\# 4;}\\
&\mathtt{!;\#0;\#0;\herhaal 15}\\
&\rris.
\end{align*}
Observe that this result implies that negative test instructions 
and basic instructions
do not increase expressiveness; indeed their sole purpose is to provide
ease of specification. On the other hand, jump instructions with counters of 
unbounded size are crucial for the above-mentioned expressiveness 
result (cf.~\cite{BL02,PZ06}). 

In \cite{BP09} we introduced an alternative for \IS:
the set \PI\ of primitive instructions that underlies PGA and \IS\
is replaced by a set of 
programming instructions that specifically prescribe whether the next 
instruction to be executed is concatenated to the right or 
to the left. The resulting semigroup C also 
produces all periodic SPIs over \PI. More results on C are discussed in~\cite{YY}.
We also mention here~\cite{BM12},
in which SPISA is extensively introduced (Single Pass Instruction Sequence Algebra),
a variant of PGA that comprises next to the termination instruction also
a positive termination instruction !t and a negative termination instruction !f.

\subsection{PGLA}
\label{YY}
The program notation PGLA, which is in fact
$\IS$ as defined in this paper, was introduced in~\cite{BL02}
as a first
example of a `programming language'.
The criterion formulated in~\cite{BL02} to use this
terminology is the existence of a projection function
\[\lata\] 
(PGLA to PGA) that maps any PGLA-program ($\IS$-expression) 
to a PGA-program. In fact, PGLA inspired a toolset and programming 
environment for program algebra \cite{Diertens}. 
However, we now
conclude that we did not deal in a proper way with the 
non-standard
case of programs with repeaters with a counter that is too
large:
the projection function $\lata$ then adds $\#0$-instructions
to obtain a first canonical \IS-form. This solution
does not combine in an elegant way
with jumps as witnessed by the following examples, where 
we write $|X|$ for the thread 
extraction of PGA-program $X$ and use the following abbreviations: 
\begin{align*}
|Y|_{pgla}~&\text{ for $|\lata(Y)|$ (as is done in \cite{BL02}), and}\\
\mathtt a^\infty~~~~~&\text{ for the thread defined by $P=\mathtt a\circ P$.}
\end{align*}
Some examples:
\begin{align*}
|\mathtt a;\#1;\herhaal 3|_{pgla}
&=|\mathtt a;\#1;\#0;\herhaal 3|_{pgla}
=|(\mathtt a;\#1;\#0)^\omega|=\mathtt a\circ\di,\\
|\mathtt a;\#2;\herhaal 3|_{pgla}
&=|\mathtt a;\#2;\#0;\herhaal 3|_{pgla}
=|(\mathtt a;\#2;\#0)^\omega|=\mathtt a^\infty,
\end{align*}
and, more generally, for $k\in\Nat$ we find
\[|\mathtt a;\#k;\herhaal 3|_{pgla}=
|(\mathtt a;\#k;\#0)^\omega|=\begin{cases}
\mathtt a^\infty&
\text{if $k\bmod3=2$},\\
\mathtt a\circ\di&\text{otherwise.}
\end{cases}\]
So in PGLA's projection of $|\mathtt a;\#k;\herhaal 3|_{pgla}$, 
deadlock $\di$ \emph{either} arises from
the added jump instruction $\#0$,
\emph{or} from the interplay with $\herhaal 3$ and the
original jump instruction $\#k$, \emph{or} does not arise. 
This we now consider rather
arbitrary and we prefer to view $\ISP$, a proper subset
of PGLA, as the program notation that is closest to PGA.

Thus, our final conclusion is to avoid the question of ``too large 
repeat counters'' and to state
that $u_1;\ldots;u_k;\herhaal n$ is not a program whenever $k>n$.
This agrees with the point of view to
consider PGA the more basic theory for providing
semantics for sequential programming from a \emph{mathematical} 
point of view (instead of $\IS$ or $\ISP$)
 and
with the point of departure adopted in~\cite{BL02}:
a programming language is a pair $(E,\phi)$ with $E$ a set 
of expressions (the programs) and $\phi$ a 
projection function to PGA.
Finally, we note that $\ISP$ shares
a property that is often seen in imperative programming:
if 
\[P;Q\]
is a $\IS$-program, then $P$ and $Q$ need not be $\IS$-programs
(while \IS-expressions satisfy by definition the property that
their decomposition yields \IS-expressions, and the same can be 
said for SPIs).

\addcontentsline{toc}{section}{References}

\bibliographystyle{plain}

\end{document}